\documentclass[fleqn,usenatbib]{mnras}
\usepackage{booktabs, siunitx, amssymb, multirow}
\usepackage[T1]{fontenc}

\DeclareRobustCommand{\VAN}[3]{#2}
\let\VANthebibliography\thebibliography
\def\thebibliography{\DeclareRobustCommand{\VAN}[3]{##3}\VANthebibliography}

\usepackage{graphicx}	
\usepackage{amsmath}	
\usepackage{amssymb}	

\title[Fragmentation and composition tracking in REBOUND]{Collisional fragmentation and bulk composition tracking in REBOUND}

\author[A. C. Childs \& J. H. Steffen]{
Anna C. Childs\thanks{E-mail: childsa6@unlv.nevada.edu}
and Jason H. Steffen
\\
$^{1}$Department of Physics and Astronomy, University of Nevada, Las Vegas, 4505 South Maryland Parkway, Las Vegas, NV 89154, USA\\
}
\date{Accepted XXX. Received YYY; in original form ZZZ}

\pubyear{2021}
\usepackage{newtxtext,newtxmath}

\begin{document}
\label{firstpage}
\pagerange{\pageref{firstpage}--\pageref{lastpage}}
\maketitle

\begin{abstract}
We present a fragmentation module and a composition tracking code for the $n$-body code \textsc{rebound}.  Our fragmentation code utilises previous semi-analytic models and follows an implementation method similar to fragmentation for the $n$-body code \textsc{mercury}.  In our $n$-body simulations with fragmentation, we decrease the collision and planet formation timescales by inflating the particle radii by an expansion factor $f$ and experiment with various values of $f$ to understand how expansion factors affect the collision history and final planetary system.  As the expansion factor increases, so do the rate of mergers which produces planetary systems with more planets and planets at larger orbits. Additionally, we present a composition tracking code which follows the compositional change of homogeneous bodies as a function of mass exchange and use it to study how fragmentation and the use of an expansion factor affects volatile delivery to the inner terrestrial disc.  We find that fragmentation enhances radial mixing relative to perfect merging and that on average, as $f$ increases so does the average water mass fraction of the planets.  Radial mixing decreases with increasing $f$ as collisions happen early on, before the bodies have time to grow to excited orbits and move away from their original location.
\end{abstract}




\section{Introduction}
The late stage of planet formation is characterised by high-energy collisions of Moon to Mars-sized bodies \citep{Weidenschilling1977, Righter11, Rafikov_2003}.  Bodies of this size are held together primarily by self-gravity and their mutual collisions are gravity dominated.  Although many of the giant impacts during this stage of planet formation will result in fragments, most $n$-body studies assume perfectly inelastic collisions, neglecting the effects of fragmentation \citep{CHAMBERS1998, AGNOR1999, CHAMBERS2001, Raymond2004, Raymond2009, Obrien2006, Morishima2010}.  Simple collision models are often used because they require less computation time than a model that includes fragmentation.  $N$-body run times depend on the number of particles being tracked.  A simple collision model that only allows inelastic collisions ensures that the number of bodies decreases with time.  However, a collision model that accounts for fragmentation can increase the number of bodies, and thus increase the run time and computation cost of the simulation.  Although accurately modeling collisions and allowing for fragments to interact with the rest of the system increases the run time, it allows for a more complex, realistic simulation that may be necessary to constrain the final physical properties of a planet.

To improve upon a simple collision model, various smooth particle hydrodynamic (SPH) studies have been conducted to understand the collision outcomes between high-velocity planetary bodies \citep{Leinhardt_2011, Leinhardt_2011b, Gabriel_2020}.  These studies provide semi-analytic models that may be incorporated into $n$-body codes to more accurately model collision outcomes.  The semi-analytic collision models of \citet{Leinhardt_2011} have been incorporated into \textsc{mercury} \citep{Chambers2013}.  Using this version of \textsc{mercury} with fragmentation, \citet{Quintana_2016} show that fragmentation does not significantly affect the multiplicity, masses and orbital elements in the final planetary system.  However, they found that fragmentation will increase the timescale for planet formation and change a planet's collision history.
\citet{Kokubo_2010} reached similar conclusions in a study modeling fragmentation.  Because previous work concluded that fragmentation does not significantly affect the final architecture of the planetary system, the importance of fragmentation in $n$-body studies has often been dismissed.  However, \citet{Emsenhuber2020} recently overturned these conclusions by showing fragmentation can profoundly alter the final planetary system by producing a greater diversity of planet sizes (although their model did not account for the re-accretion of debris). 

\textsc{rebound} is an open-source $n$-body code that is quickly becoming a standard for $n$-body problems in astronomy \citep{Rein2012}. The framework of \textsc{rebound} permits users to contribute their own modules with relative ease.  This allows \textsc{rebound} to handle various problems with up-to-date physics models.
 In this paper we introduce a new fragmentation module for \textsc{rebound}.  We implement a realistic collision model into this module based on \citet{Leinhardt_2011}.  We compare our fragmentation code in \textsc{rebound} to the fragmentation code in \textsc{mercury}.  We also present a post-processing code \textsc{rebound} which tracks how the composition of bodies with homogeneous compositions changes as they collide with one another to form planets. As planetesimals and embryos of differing composition collide to build planets there will be an exchange of mass that determines the final composition of the planet \citep{Moriarty2014}.
 
 We show an example of our codes where we examine how volatile delivery to the inner solar system is affected when fragmentation is accounted for.  Earth is considered a ``dry'' planet but it is the wettest of the terrestrial planets with $\sim 0.05-0.1\%$ of its mass being water \citep{LECUYER1998, Marty2012}.  It is theorized that Earth's water was delivered at a later stage of planet formation by wet carbonaceous chondrites from the outer asteroid belt ($2.5-4 \, \rm  au$) \citep{Morbidelli2000}.  Previous $n$-body studies, using only perfect accretion, examine how water is delivered to and accreted by the terrestrial planets of the solar system \citep{Raymond2004, Raymond2006}.  Using the initial water distribution in \citet{Raymond2006}, we use our two new codes to study how water delivery to the terrestrial planets is affected when fragmentation is considered.

In order to reduce computation time, some $n$-body studies employ an expansion factor $f$ that expands the initial particle radii by some value \citep{Leinhardt2005, Bonsor2015, Carter2015}.  The use of an expansion factor was shown by \citet{Kokubo_1996, Kokubo_2002} to not have a significant effect on the evolution of planets other than reducing the timescale of planet formation provided that the velocity dispersion of the bodies is not dominated by gravitational scattering.  While some of these studies include the effects of fragmentation, only expansion factors up to $f=6$ that also model fragmentation have been used.  In an effort to characterise the effects of an expansion factor with our fragmentation code we study how the magnitude of the expansion factor affects the final planetary system architecture, the collision history, and volatile delivery into the inner terrestrial region.

In Section 2 of this paper, we outline our fragmentation model and compare our fragmentation results in \textsc{rebound} to fragmentation results in \textsc{mercury}.  In Section 3 we show how expansion factors affect the final planetary system and collision history when fragmentation is modeled.  In section 4 we present our module for tracking the bulk composition of terrestrial planets and in Section 5 we show how volatile delivery to the inner solar system is affected by fragmentation and by different expansion factors.  In our final section we summarise our findings.  Our fragmentation and composition tracking codes are made publicly available to be used for future astrophysical studies with \textsc{rebound}.

\section{Collisional Fragmentation}

\begin{figure}
	\includegraphics[width=\columnwidth]{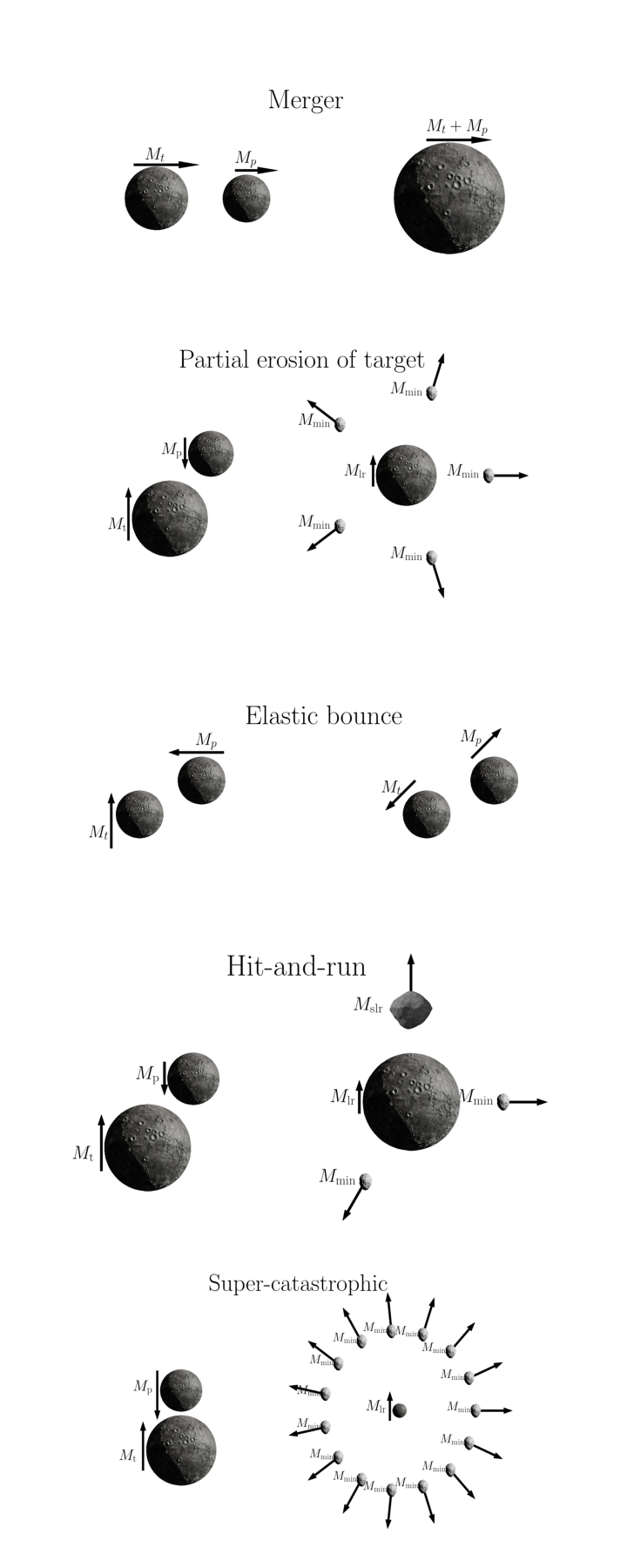}
    \caption{2-D illustrations of the various collision outcomes allowed in our fragmentation code.  For each collision outcome listed, the image on the left depicts the pre-collision geometry and the image on the right depicts the post-collision geometry.  The velocity vector and mass for each body is also shown.}
    \label{fig:collision_outcomes}
\end{figure}

Our fragmentation model for \textsc{rebound} is based on the work of \citet{Leinhardt_2011}, \citet{Asphaug2010}, and \citet{Genda2012}, and closely follows the implementation methods of \citet{Chambers2013}.  \citet{Leinhardt_2011} prescribe collision outcomes for a range of impact velocities and impact parameters of a collision. \citet{Asphaug2010} and \citet{Genda2012} refines this prescription for the subset of possible collision outcomes known as \textit{grazing impacts}.  We compare our model in \textsc{rebound} to the fragmentation model written by \citet{Chambers2013} for the $n$-body code \textsc{mercury}, whose fragmentation model is also based on the work of \citet{Leinhardt_2011, Asphaug2010, Genda2012}.  We extend on previous work and allow for bodies to have different densities.  If the target and projectile remain in the system after a collision, their pre-collision densities are maintained regardless of any mass exchange.  If any fragments are produced, they are assigned the density of the target.

Our fragmentation model is written in \textsc{C} and is designed to work with \textsc{rebound}.  It may be easily updated as more accurate collision models become available.  In this section we present the fragmentation model, the setup of our $n$-body simulations, and our comparative results.

\subsection{Fragmentation model}\label{sec:Fragmentation_model}
 \textsc{rebound} offers several collision detection modules.  We test our code using the ``direct detection" module which checks for overlaps of particle radii between every particle pair at every time step. However, regardless of the collision detection module chosen, in a system with $n$ non-zero mass particles with a non-zero radii, a collision is detected at the time when the physical radii of two bodies come into contact. Since this criteria must be met for a collision, we expect our fragmentation code to be compatible with all the collision detection modules in \textsc{rebound}.
 
 The more massive of the two bodies is labeled the \textit{target} with mass $M_{\rm t}$ and radius $R_{\rm t}$, and the other body is labeled the \textit{projectile} with mass $M_{\rm p}$ and radius $R_{\rm p}$.  To resolve a collision, we first find the impact velocity, $v_{\rm i}$, and impact parameter, $b$.  The impact velocity, which follows from orbital energy conservation, is 
\begin{equation}
    v_{\rm i} = \sqrt{{v_{\rm rel}}^2-2GM_{\rm tot}( \frac{1}{x_{\rm rel}} - \frac{1}{R_{\rm tot}})},
\end{equation}
where $v_{\rm rel}$ is the relative velocity between the two bodies, $G$ is the gravitational constant, $R_{\rm tot}$ is the sum of the target and projectile radii, and $x_{\rm rel}$ is the distance between the two centers of the bodies.

The impact angle, $\theta$, is  the angle between the line connecting the centers of the two bodies and their relative velocity vector at the time of the collision.  The impact parameter, $b$, is the distance between the two centers and is projected perpendicular to the impact velocity vector,
\begin{equation}
 b= R_{\rm tot}\textrm{sin} \theta\ =\frac{\mid\textbf{h}\mid}{v_{\rm i}},
\end{equation}
where \textbf{h} is the angular momentum vector of the colliding bodies. The collision geometry for the parameters $v_{\rm i}$ and $b$ may be seen in Figure 2 of \citet{Leinhardt_2011}.

\subsubsection{Perfect merging}
 The mutual escape velocity of the colliding system is given by
\begin{equation}
    v_{ \rm esc}= \sqrt{2GM_{\rm tot}/R_{\rm tot}}.
\end{equation}
Regardless of impact parameter, if the impact velocity of the collision is less than or equal to the mutual escape velocity of the two bodies the collision results in a perfect merger.  A perfect merger is modeled in the same manner as an inelastic collision where the collision results in one body and mass and momentum are conserved.  The projectile is merged into the target, thus removing the projectile from the simulation.

\subsubsection{Partial and super-catastrophic erosion of the target}
In all other collision scenarios, the impact velocity will be greater than the mutual escape velocity and the collision is resolved as a function of $b$ and the mass of the largest post-collision body, $M_{\rm lr}$.  $M_{\rm lr}$ is the largest remnant to survive a collision and is found with
\begin{equation}\label{eq:universal_law}
    M_{\rm lr}=M_{\rm tot} \left (1 - \frac{Q}{2Q^{'}}  \right ),
\end{equation}
if $Q/{Q^{'}} > 1.8$.  Otherwise,
\begin{equation}
    M_{\rm lr}=0.1M_{\rm tot} \left (\frac{Q}{1.8Q^{'}} \right )^{-3/2},
\end{equation}
where $Q$ is the center of mass specific impact energy given by
\begin{equation}
    Q = \frac{\mu v_{\rm i}^{2}}{2M_{\rm tot}},
\end{equation}
and $\mu$ is the reduced mass, $M_{\rm t}M_{\rm p}/M_{\rm tot}$. This relationship is derived empirically from \citep{Stewart_2009}.

The catastrophic disruption threshold is the energy required to disperse half of the total mass in a collision and is defined as
\begin{equation}
    Q^{'}=Q_{0}^{'}=0.8c^{*} \pi \rho_{1} G R_{1}^2
\end{equation}
for head-on ($b=0$) collisions of equal sized bodies.  $c^{*}$ is a dimensionless material parameter that represents the offset between the gravitational binding energy and the catastrophic disruption threshold for impact energy.  For terrestrial planetary bodies, which are used in this study, $c^{*}=1.8$.  $R_{1}$ is the combined radius of the bodies with an assumed density of $\rho_{1} \equiv 1000$ ${\rm kg}$ ${\rm m}^{-3}$.

If the collision has $b>0$ or involves bodies of different mass, the catastrophic disruption threshold is more generally defined as
\begin{equation}
    Q^{'}=Q_{0}^{'} \left( \frac{(1+\gamma)^2}{4 \gamma} \right ) \left ( \frac{\mu}{\mu_{\alpha}} \right)^{3/2}
\end{equation}
where $\gamma=M_{\rm p} / M_{\rm t}$ and $\mu_{\alpha}$ is the reduced mass for the fraction $\alpha$ that is involved in the collision,
\begin{equation}
    \alpha \equiv \frac{3R_{\rm p}l^2-l^3}{4R_{\rm p}^{3}}
\end{equation}
with $l=R_{\rm tot}(1 -{\rm sin}\theta)$ and
\begin{equation}
    \mu_{\alpha}=\frac{\alpha M_{\rm p} M_{\rm t}}{M_{\rm t}+\alpha M_{\rm p}}.
\end{equation}

The largest post-collision body is assigned the mass $M_{\rm lr}$ and is placed at the center of mass (at the time of the collision) position and velocity.  If $M_{\rm lr}<M_{\rm t}$ then partial erosion of the target will take place.  The remaining mass, $M_{\rm r}=M_{\rm tot}-M_{\rm lr}$, is broken up into equal sized fragments.  The number of fragments, $N_{\rm frag}$, made in the collision is given by
\begin{equation}
    N_{\rm frag} = \left\lfloor\dfrac{M_{\rm r}}{M_{\rm min}}\right\rfloor
\end{equation}
where $M_{\rm min}$ is the minimum fragment mass defined by the user. The fragment mass is then $M_{\rm r}/N_{\rm frag}$.  The fragments are uniformly placed in a circle in the collision plane at an equal distance from the center of mass and given an initial velocity of $1.1 v_{\rm esc}$.  The collision plane is defined by the the line connecting the centers of the bodies and the impact velocity vector.  Velocities and positions of the resulting bodies are then adjusted to conserve momentum. The geometric distribution of the fragments is shown in Figure \ref{fig:collision_outcomes}.  Assigning the fragments a velocity of  $1.1 v_{\rm esc}$ is likely an underestimate of the true fragment velocity, which may artificially dampen the dynamical state of the system. The geometry and velocity of the post-collision remnants is a somewhat arbitrary choice in \citet{Chambers2013}, but we adopt these same choices in our code so that we may make a direct comparison between the two fragmentation codes.  However, our code allows for the user to change these distributions with relative ease as more accurate models for geometric and velocity distributions of the post-collision fragments become available.

If $M_{\rm lr} \leq 0.1 M_{\rm t}$, the collision is referred to as a super-catastrophic collision.  If the remaining mass from a collision that erodes the target is less than $M_{\rm min}$ then the collision results in an effective merger, where the outcome is the same as in the case of a perfect merger.

\subsubsection{Grazing impacts}
If $b \geq b_{\rm crit}$, the collision is defined as a grazing impact and multiple collision outcomes are possible.  Following \citet{Asphaug2010},
\begin{equation}
    b_{\rm crit}= R_{\rm t},
\end{equation}
and in addition to partial erosion of the target as discussed above, a grazing impact may also result in a graze-and-merge, a hit-and-run, or an elastic bounce.  A graze-and-merge event happens when a grazing impact results in fragments whose velocity will be less than the escape velocity of the original system.  As a result, the fragments are re-accreted by the target and the result is that of a perfect merger.  \citet{Genda2012} defined the criteria for a graze-and-merge as $v_{\rm i} \leq v_{\rm cr}$ where
\begin{equation}
    v_{\rm cr}= v_{\rm esc} \left [ c_{1} \Gamma (1 -\rm{sin} \theta)^{5/2} + c_{2} \Gamma + c_{3} \Gamma (1-\rm{sin} \theta)^{5/2} + c_{4}  \right ],
\end{equation}

\begin{equation}
    \Gamma = \left ( \frac{1- \gamma}{1+ \gamma} \right ) ^ {2},
\end{equation}
and
\begin{equation}
    c_{1}=2.43,  \:
   c_{2}=-0.0408,\:
  c_{3}=1.86 \: \textrm{and} \:
  c_{4}=1.08.
\end{equation}

Unlike the fragmentation code developed by \citet{Chambers2013}, our model allows for partial accretion of the projectile by the target.  If $M_{\rm lr} \geq M_{\rm t}$ then the target is assigned the mass of $M_{\rm lr}$ and the remaining projectile mass is partitioned into fragments.  In the more specific instance where a grazing impact takes place, $v_{\rm i} > v_{\rm cr}$ and $M_{\rm lr} \geq M_{\rm t}$, then the collision is defined as a hit-and-run.  The target is assigned the $M_{\rm lr}$ and the mass of the largest remnant for the projectile, or the second largest remnant $M_{\rm slr}$, must be calculated.  The process for doing so is similar to calculating the mass of the largest remnant as described above, however the roles of the target and projectile are switched as we now consider the reverse impact.  The center of mass specific impact energy is now given by
\begin{equation}
    Q = \frac{\mu v_{\rm i}^2}{2(\beta M_{\rm t}+M_{\rm p})}
\end{equation}
where 
\begin{equation}
    \mu = \frac{\beta M_{\rm t} M_{\rm p}}{\beta M_{\rm t} + M_{\rm p}}
\end{equation}
and $\beta$ is the fraction of the target that would intersect the projectile in a head-on collision. The critical value, $Q'$, for the reverse impact is
\begin{equation}
    Q'= \left [ \frac{(\gamma+1)^2}{4 \gamma}  \right]Q_{0}',
\end{equation}
\begin{equation}
    \gamma = \frac{\beta M_{\rm t}}{M_{\rm p}},
\end{equation}
and $Q_{0}'$ is defined the same way as before.

The mass of the largest remnant of the projectile is then
\begin{equation}
\begin{array}{l}
    M_{\rm slr} = (\beta M_{\rm t} + M_{\rm p}) \left( 1 - \frac{Q}{2Q'} \right) \qquad Q<1.8Q' \\
    \\
    \qquad = 0.1(\beta M_{\rm t} + M_{\rm p})\left(\frac{Q}{1.8Q'}\right)^{-1.5} \qquad Q \geq 1.8Q'.
    \end{array}
\end{equation}
If $M_{\rm p} - M_{\rm slr} < M_{\rm min}$, then the projectile is assigned its initial mass and an elastic bounce takes place.

Figure \ref{fig:collision_outcomes} depicts the pre-collision and post-collision geometry for the different types of collisions allowed in our code. A summary of collision outcomes as a function of impact parameter and velocity is given in Table \ref{tab:collision_types}.

\begin{table} 
\centering
\setlength{\tabcolsep}{5.0pt}     
\setlength{\cmidrulekern}{0.3em} 
\caption{Collision outcomes as a function of impact parameter, $b$, impact velocity, $v_{\rm i}$, and the post-collision remnant masses.}
\begin{tabular}{%
  S[table-format=1.0]
  S[table-format=1.2]
  *{2}{ 
    *{2}{S[table-format=1.3(3)]} 
    S[table-format=1.3]
  }
}
  \toprule

 {Collision outcome} & {$b$} & {$v_{\rm i}$} & {Mass constraints} \\ 
 \cmidrule(lr){1-4} 
 {Perfect merger} & {all} & {$\leq v_{\rm esc}$}& {-} \\ 
 {Effective merger} & {all} & {$> v_{\rm esc}$} & {$M_{\rm tot}-M_{\rm lr} < M_{\rm min}$}\\ 
  {Graze-and-merge} & {$\geq b_{\rm crit}$} & {$\leq v_{\rm crit}$} & {-}\\ 
  {Partial erosion of target} & {all} & {$> v_{\rm esc}$} & {$M_{\rm lr} < M_{t}$}\\ 
  {Elastic bounce} & {$\geq b_{\rm crit}$} & {$> v_{\rm crit}$} & {$M_{\rm p} - M_{\rm slr} < M_{\rm min}$}\\
    {Partial accretion by target} & {all} & {$> v_{\rm esc}$} & {$M_{\rm lr} > M_{t}$}\\ 

{Hit-and-run} & {$\geq b_{\rm crit}$} & {$> v_{\rm crit}$} & {$M_{\rm slr} < M_{p}, M_{\rm lr} \geq M_{t}$}\\ 
  {Super-catastrophic} & {all} & {$> v_{\rm esc}$} & {$M_{\rm lr} \leq 0.1M_{t}$}\\ 

  \bottomrule
\end{tabular}
\label{tab:collision_types}
\end{table}

\subsection{Comparison Setup}\label{sec:n_body_setup}

\begin{figure*}
	\includegraphics[width=1.5\columnwidth]{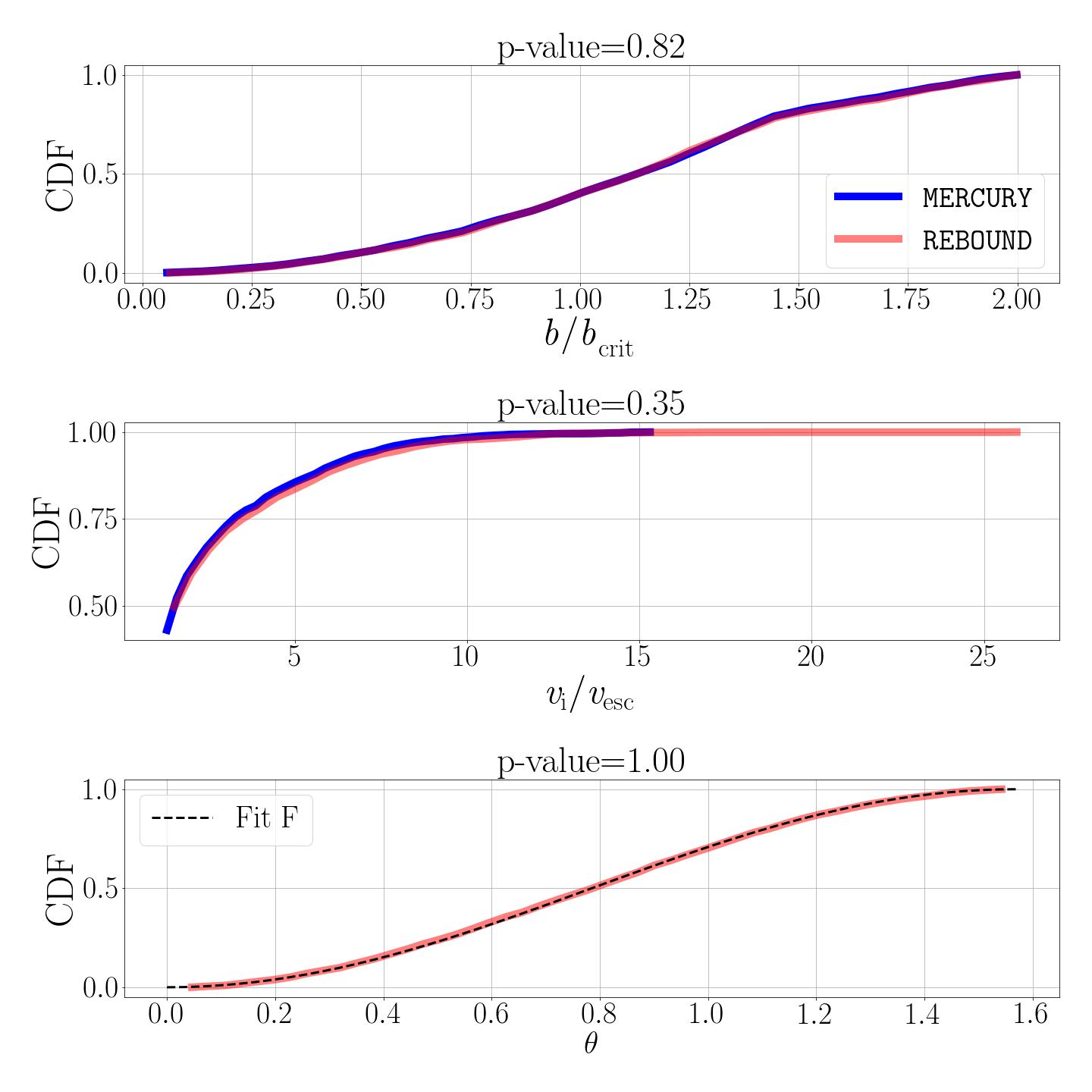}
    \caption{Cumulative distribution functions (CDF) for the impact parameters (top panel) and impact velocities (middle panel) for all the collisions in each $n$-body code after $5 \, \rm Myr$ of integration time. We perform a two-sample KS-test and report the p-values.  The large p-values indicate that the data are consistent with being drawn from the same distributions and the two $n$-body codes are producing similar collisions.  \textsc{rebound} returns one outlier, a collision between two planetesimals with $v{\rm i}/v_{\rm esc}$ > 25.  In the bottom panel we show the CDF of the impact angles ($\theta$) in \textsc{rebound} and the expected CDF fit, Fit F, from \citet{Shoemaker1962}.}
    \label{fig:params}
\end{figure*} 

We simulate the late stage of planet formation and compare our results in \textsc{rebound} to the results of \citet{Chambers2013} fragmentation model in \textsc{mercury}.  Both codes use the same starting discs and initial conditions unless otherwise stated.  

We adopt a disc of small planetesimals and larger planetary embryos from \citet{CHAMBERS2010}, which successfully reproduces the broad characteristics of the solar system's terrestrial planets.  This bimodal mass distribution marks the epoch of planet formation which is dominated by purely gravitational collisions in which our fragmentation model is valid \citep{KOKUBO2000}.  This disc has also been used in other studies of the solar system \citep{Quintana_2016, Childs_2019}.  Our disc contains 26 embryos (Mars-sized, $r = 0.56 R_\oplus$; $m=0.093 M_\oplus$), and 260 planetesimals (Moon-sized, $r=0.26 R_\oplus$; $m=0.0093 M_\oplus$) yielding a total disc mass of 4.85 $M_{\oplus}$.  The minimum fragment mass is set to half the mass of a planetesimal, $M_{\rm min}=0.0047 M_\oplus$.  The disc has no gas.  All masses have a uniform density of $3 \, \rm {g \, cm}^{-3}$.  The surface density distribution, $\Sigma$, of the planetesimals and embryos follows $\Sigma \sim r^{-3/2}$, the estimated surface density distribution of Solar Nebula models \citep{Weidenschilling1977}.  The masses are distributed between $0.35 \, \rm au$ and $4.0 \, \rm au$ from a Sun-like star.

The eccentricities and inclinations for each body are drawn from a uniform distribution of $e< 0.01$ and $i< 1^{\circ}$.  The argument of periastron ($\omega$), mean anomaly ($M$), and longitude of ascending node ($\Omega$) are chosen at random from a uniform distribution between $0^{\circ}$ and $360^{\circ}$.  The random generator seed is changed in each run to provide variation in the distribution of these orbital elements for all the bodies in \textsc{rebound}.  In \textsc{mercury}, an initial disk is generated with the same distributions mentioned above and the orbital elements of one planetesimal are chosen randomly in each run to provide variation between the runs.  We include Saturn with $m=95.1 \, M_{\oplus}$, $a=9.54309\, \rm au$, $e=0.052519$, $i=0.8892^{\circ}$, $\Omega=90^{\circ}$, $\omega=324.5263^{\circ}$, and $M=256.9188^{\circ}$.  We include Jupiter with $m=317.7 \, M_{\oplus}$, $a=5.20349\, \rm au$, $e=0.048381$, $i=0.3650^{\circ}$, $\Omega=0.0^{\circ}$, $\omega=68.3155^{\circ}$, and $M=227.0537^{\circ}$.

We use the hybrid integrator available in each $n$-body code. In \textsc{rebound} the integrator \textsc{Mercurius} is used.  It is a hybrid of the symplectic \textsc{WHFast} integrator and the non-symplectic \textsc{IAS15} integrator \citep{Rein2019}.  In \textsc{mercury}, the integrator used is a hybrid of a second-order mixed-variable symplectic and \textsc{Bulirsch-Stoer} integrator \citep{Chambers1999}.  We choose initial time-steps of six days, which is $\frac{1}{10}$ of the inner-most orbit.  40 runs are performed with each $n$-body code and each run is integrated for $5 \, \rm Myr$.  We note that while we are using hybrid integrators in both $n$-body codes, the integrators are different and will contribute different numerical effects to the $n$-body results.

\subsection{Comparison results}
\begin{figure*}
	\includegraphics[width=2\columnwidth]{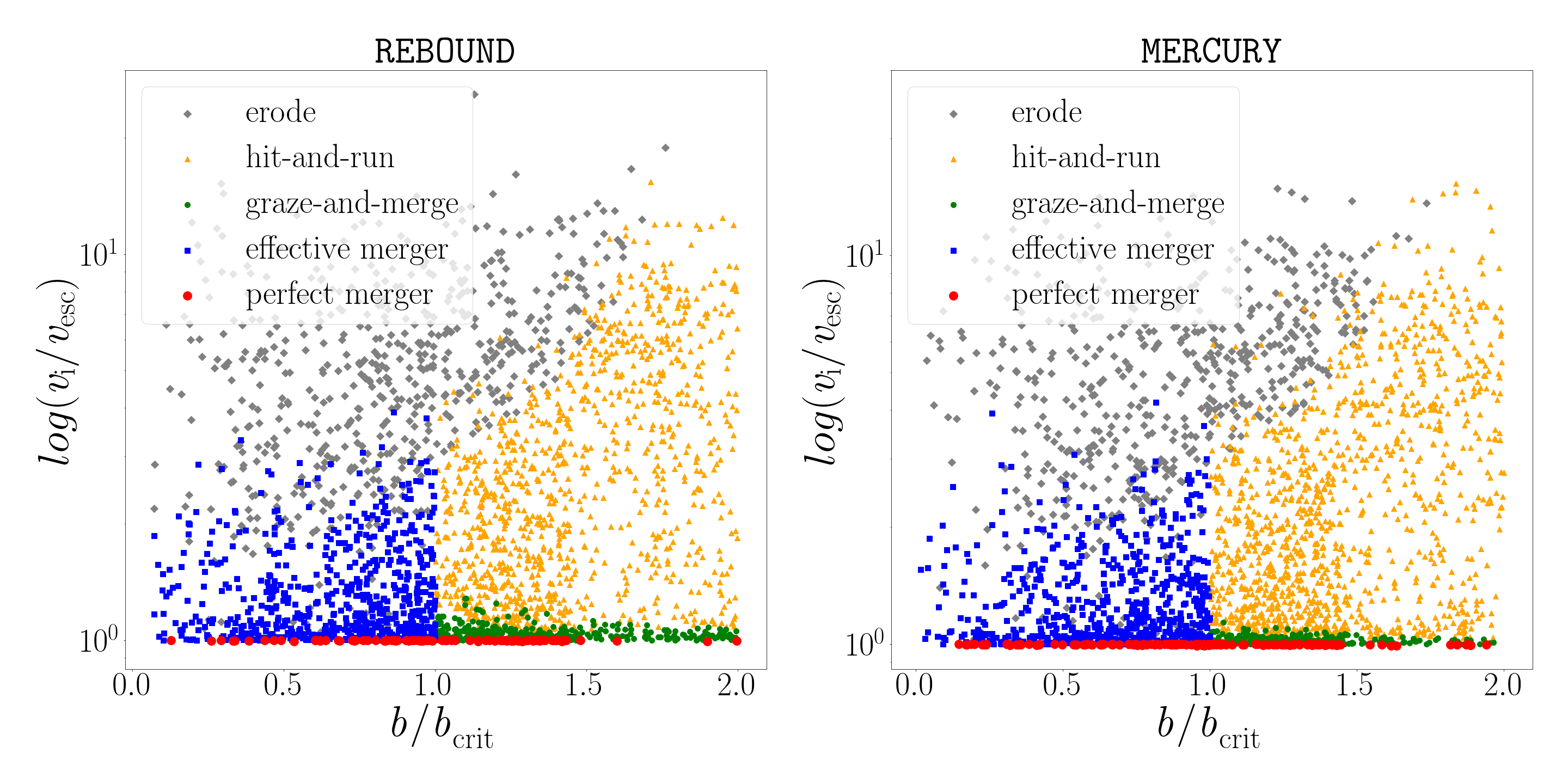}
    \caption{Collision outcomes as a function of impact parameter and impact velocity in both \textsc{rebound} and \textsc{mercury}.  Collisions that occur in the same parameter space are resolved in the same manner by both codes.}
    \label{fig:collision_types}
\end{figure*}

Because the most important parameters in determining the outcome of a collision is the impact velocity, $v_i$, and the impact parameter, $b$, we first consider the distribution of these two parameters in Figure \ref{fig:params}.  In the top panel we show the cumulative distribution function (CDF) for $b/b_{\rm crit}$ and in the middle panel we show the CDF for $v_{\rm i}/v_{\rm esc}$ for all the collisions in both $n$-body codes.  We perform a two-sample KS-test for each parameter.  The p-value for $b/b_{\rm crit}$ is 0.82 and the p-value for $v_{\rm i}/v_{\rm esc}$ is 0.35.  By assuming that a p-value $>0.10$ implies that we can not reject the null hypothesis, that both set are drawn from the same distribution, we conclude that \textsc{rebound} and \textsc{mercury} are producing the same kinds of collisions.  We observe a high velocity impact for one collision between two planetesimals in \textsc{rebound}, such that $v_{\rm i}/v_{\rm esc}$ >25, but determine this is an insignificant outlier in the data.   We also check that the impact angles, $\theta$, are consistent with the expected distribution $dP \propto cos(\theta)sin(\theta)d\theta$, the cumulative distribution of which is F$=(1-cos(2\theta))/2$, from \cite{Shoemaker1962}.  We show the CDF of $\theta$ in \textsc{rebound} as well as the fit F in the bottom panel in Figure \ref{fig:ef_parameter_comparison} and find agreement between our distribution and the expected distribution with a p-value of 1.00.

\begin{figure*}
	\includegraphics[width=1.5\columnwidth]{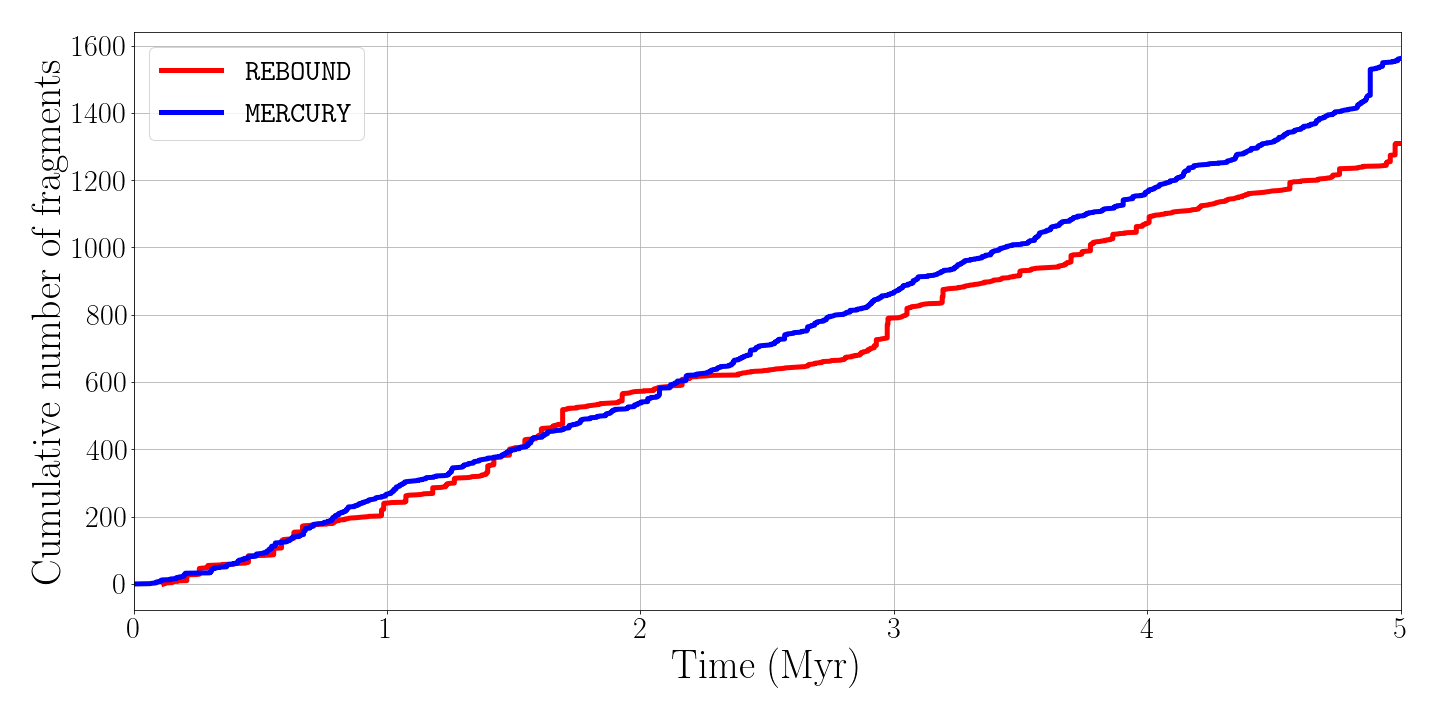}
    \caption{Total number of fragments across all 40 runs for both \textsc{rebound} and \textsc{mercury}, as a function of simulation time.  The differences in cumulative number of fragments at $5 \, \rm Myr$ can be attributed to the fact that we permit partial accretion.  In general, we observe good agreement in fragment growth between the two codes. }
    \label{fig:frag_growth}
\end{figure*}

Next, we consider the outcomes of these collisions.  Figure \ref{fig:collision_types} shows the different collision types for both codes as a function of $v_{\rm i}/v_{\rm esc}$ and $b/b_{\rm crit}$.  Different collision outcomes are denoted by a unique color and symbol.  We find that the collisions that occur in the same parameter space are resolved in the same manner in each code.  An erosion event is any event that produces fragments which yield even more collisions as the fragments are free to collide with other bodies.  Figure \ref{fig:frag_growth} compares the growth of fragments versus simulation time across all 40 runs for each code.  We see that our code produces fewer fragments after $5 \, \rm Myr$ with \textsc{mercury} producing an average of 7 more fragments than \textsc{rebound} in a run.  We attribute this difference to the fact that we allow for partial accretion in our code.  In \citet{Chambers2013}, if $M_{\rm t}<M_{\rm lr}$ the $M_{\rm lr}$ is assigned the mass of the initial target and the remaining mass is partitioned into fragments.  This results in more mass being available for the creation of fragments compared to our code where the target is allowed to partially accrete mass from the projectile.  The linearity observed in fragment growth in \textsc{mercury} is likely the result of how the discs were initialized.  The \textsc{mercury} discs do not begin as randomized as the \textsc{rebound} discs and as a result, fragment growth in \textsc{mercury} is more linear for the first $5 \, \rm Myr$. However, we observe a spike in fragments near $5 \, \rm Myr$ indicating a departure from the linear growth.  In general, fragment growth between the two codes is similar.

\section{Expansion factor}

The practice of artificially inflating particle radii by an expansion factor $f$ is sometimes used in $n$-body studies to reduce the computation time \citep{Kokubo_2002, Carter2015, Childs2021}.  Expanding the particle radii by $f$ is equivalent to decreasing the density of the body by $1/f^3$, thereby increasing the collision probability.  Increasing the collision rate speeds up planet formation and decreases computation time and has been shown not to affect $n$-body outcomes significantly for $f \leq 10$ \citep{Kokubo_2002, Carter2015}.  \citet{Childs2021} experimented with expansion factors $f=25$ and $f=50$ while using only perfect merging and found that the larger the expansion factor is, the more quickly the bodies merge together.  This leads to systems with higher planet multiplicities and damped orbits since the planets form before the orbits have time to reach excited states via planet-planet interactions.

Here, we are interested in understanding how expansion factors affect $n$-body results when fragmentation is accounted for.  In order to decrease computation time, the smallest expansion factor we experiment with is $f=3$, which we take as our fiducial system.  It takes two weeks to integrate the $f=3$ systems for 100 Myr whereas it takes one and a half weeks to integrate the $f=1$ system for 5 Myr and we need to integrate the $f=1$ system for hundreds of Myrs for the systems to fully evolve.  We experiment with $f=3,5,7,10,15,20,35, \rm and \, 50$ to understand how the magnitude of the expansion factor changes the final system architecture and collision history.  For each expansion factor we conduct 10 runs using the same setup as described in Section \ref{sec:n_body_setup} and our fragmentation code for \textsc{rebound}. 

To compare different values of $f$ we must use a standard metric other than time since different values of $f$ speed up planet formation at different rates.  We choose to evaluate formation rates as a function of the remaining disc mass, mass that has not been ejected or accreted by planets, $M_{\rm d}$.   $M_{\rm d}$ is the cumulative mass of the bodies with a mass smaller than $0.1 \, M_{\oplus}$ at the end of the simulation.  The remaining disc mass is a proxy for how far along the system is in its evolution.  $0.05 \, M_{\oplus}$ is the average $M_{\rm d}$ for the $f=3$ systems after $100 \, \rm Myr$ of integration time. Because the $f=3$ systems evolve the slowest and are the most computationally restrictive, we integrate the systems with $f>3$ to a time in which they have an average $M_{\rm d} \leq 0.05 \, M_{\oplus}$.   Generally, as the expansion factor increases we integrate for less time. 

\begin{table*}
\centering
\caption{Expansion factors ($f$), simulation time and the average values and standard deviations for the final terrestrial planet multiplicity, planet mass ($M_{\rm p}$), semi-major axis ($a_{\rm p}$), eccentricity ($e$) and inclination ($i$) across all ten runs.  These statistics only consider bodies with a mass larger or equal to $0.1 \, M_{\oplus}$ and the data from 10 runs for each setup.  We also include the average remaining disc mass $M_{\rm d}$ of the runs for each $f$.}
\begin{tabular}
{%
  S[table-format=1.0]
  S[table-format=1.2]
  *{2}{ 
    *{2}{S[table-format=1.3(3)]} 
    S[table-format=1.3] 
     S[table-format=1.2]
  }}
  \toprule
    {$f$} & {Time/Myr} &{No. of planets} & {$M_{\rm p}/M_\oplus$} & {$a_{\rm p}/ \rm au$} & {$e_{\rm p}$} & {$i^{\circ}_{\rm p}$} & {$M_{\rm d}/M_\oplus$}\\
    \cmidrule(lr){1-8}   
    {3}&{100} & {$4.1 \pm 0.87$} & {$0.67 \pm 0.87$} & {$1.05 \pm 0.59$} & {$0.07 \pm 0.07$} & {$1.76 \pm 1.5$} & {$0.05 \pm 0.03$}\\ 
    
    {5}&{10}& {$5.5 \pm 1.5$} & {$0.49 \pm 0.33$} & {$1.41 \pm 0.82$} & {$0.08 \pm 0.09$} & {$2.26 \pm 1.71$}& {$0.006 \pm 0.005$}\\
    
    {7}&{10}& {$5.3 \pm 0.78$} & {$0.60 \pm 0.34$} & {$1.35 \pm 0.79$} & {$0.09 \pm 0.08$} & {$1.99 \pm 1.21$}& {$0.006 \pm 0.007$}\\  
    
    {10}&{5}&{$4.9 \pm 1.4$} & {$0.66 \pm 0.53$} & {$1.42 \pm 0.88$} & {$0.08 \pm 0.07$} & {$2.54 \pm 1.75$}& {$0.007 \pm 0.009$}\\ 
    
    {15}&{5}&{$5.8 \pm 1.5$} & {$0.63 \pm 0.44$} & {$1.34 \pm 0.74$} & {$0.06 \pm 0.03$} & {$2.81 \pm 1.76$}& {$0.007 \pm 0.009$}\\ 
    
    {20}&{5}&{$6.2 \pm 1.66$} & {$0.62 \pm 0.53$} & {$1.36 \pm 0.86$} & {$0.06 \pm 0.06$} & {$3.36 \pm 2.07$}& {$0.008 \pm 0.01$}\\ 
    
    {35}&{5}&{$6.7 \pm 1.5$} & {$0.63 \pm 0.43$} & {$1.46 \pm 0.88$} & {$0.07 \pm 0.05$} & {$2.99 \pm 2.18$}& {$0.01 \pm 0.02$}\\
    
    {50}&{5}&{$6.2 \pm 0.75$} & {$0.70 \pm 0.50$} & {$1.44 \pm 0.88$} & {$0.06 \pm 0.04$} & {$2.50 \pm 1.95$}& {$0.01 \pm 0.02$}\\ 
    
  \bottomrule
\end{tabular}
\label{tab:expansion_factors}
\end{table*}

\subsection{Effects on system architecture}

Table \ref{tab:expansion_factors} lists the expansion factor used, the time the runs were integrated for, and the average values $\pm$ the standard deviation between runs for the planet multiplicity, mass ($M_{\rm p}$), semi-major axis ($a_{\rm d}$), eccentricity ($e_{\rm p}$), and inclination ($i^{\circ}_{\rm p}$), and average remaining disc mass ($M_{\rm d}$) of the runs. We define a planet as a body having a mass of $\geq 0.1 M_{\oplus}$.

We find that as the expansion factor increases the planet multiplicity increases and the average semi-major axis that fragments are formed at also increases.  The increase in planet multiplicity happens because collisions happen on shorter timescales allowing the bodies to grow more quickly.  The shorter collision timescale also allows planets to form at larger radii and accrete nearby planetesimals, inhibiting the inward scattering of planetesimals.  There is no noticeable correlation between $f$ and average planet eccentricity, inclination, and mass.

\begin{table*}
\centering
\caption{We list the simulation time, total number of collisions, the percent of collision outcomes (mergers, elastic bounces, hit-and-runs and erosion events), and the total number of ejections, star collisions and fragments across all ten runs for each expansion factor $f$.}
\begin{tabular}
{%
  S[table-format=1.0]
  S[table-format=1.2]
  *{2}{ 
    *{2}{S[table-format=1.3(3)]} 
    S[table-format=1.3] 
     S[table-format=1.2]
  }
}
  \toprule
    {$f$} & {Time/Myr} & {Total $\#$} & {Merge (\%)} & {Elastic bounce (\%)} & {Hit-and-run (\%)} & {Erosion (\%)} & {$\#$ of ejections} & {$\#$ of star collisions} & {$\#$ of fragments}\\
    \cmidrule(lr){1-10}   
    {$3$}&{100}& {12131} & {42.1} & {35.1} & {1.4} & {21.3} & {1015} & {753} & {6125}\\
    {$5$}&{10}& {10121} & {44.5} & {37.2} & {1.8} & {16.4} & {832} & {364} & {5096}\\
    {$7$}&{10}& {11788} & {44.5} & {38.7} & {2.2} & {14.7} & {797} & {346} & {5484}\\
    {$10$}&{5}& {12383} & {49.9} & {36.9} & {2.9} & {10.3} & {652} & {145} & {6128}\\
    {$15$}&{5}& {12476} & {52.7} & {37.4} & {3.1} & {6.8} & {631} & {103} & {5882}\\
    {$20$}&{5}& {10640} & {57.2} & {35.3} & {3.4} & {4.1} & {449} & {41} & {4632}\\
    {$35$}&{5}& {12527} & {55.7} & {37.7} & {4.2} & {2.4} & {292} & {17} & {5323}\\
    {$50$}&{5}& {11323} & {56.5} & {38.0} & {4.1} & {1.3} & {142} & {2} & {4375}\\

  \bottomrule
\end{tabular}
\label{tab:ef_collision_history}
\end{table*}

\subsection{Effects on collision history}

Table \ref{tab:ef_collision_history} lists the total number of collisions, the percentages of collision outcomes, and the total number of ejections, star collisions and fragments for all ten runs with a given expansion factor.  We observe a correlation between $f$ and the occurrence rate of some of the collision outcomes -- as the expansion factor increases, the fraction of mergers, and hit-and-run collisions increases, but the fraction of erosion events, the number of ejections, and star collisions decrease.  We observe no noticeable correlation between expansion factor and the total number of collisions and fragments, or the fraction of elastic bounces.

\begin{figure*}
	\includegraphics[width=2\columnwidth]{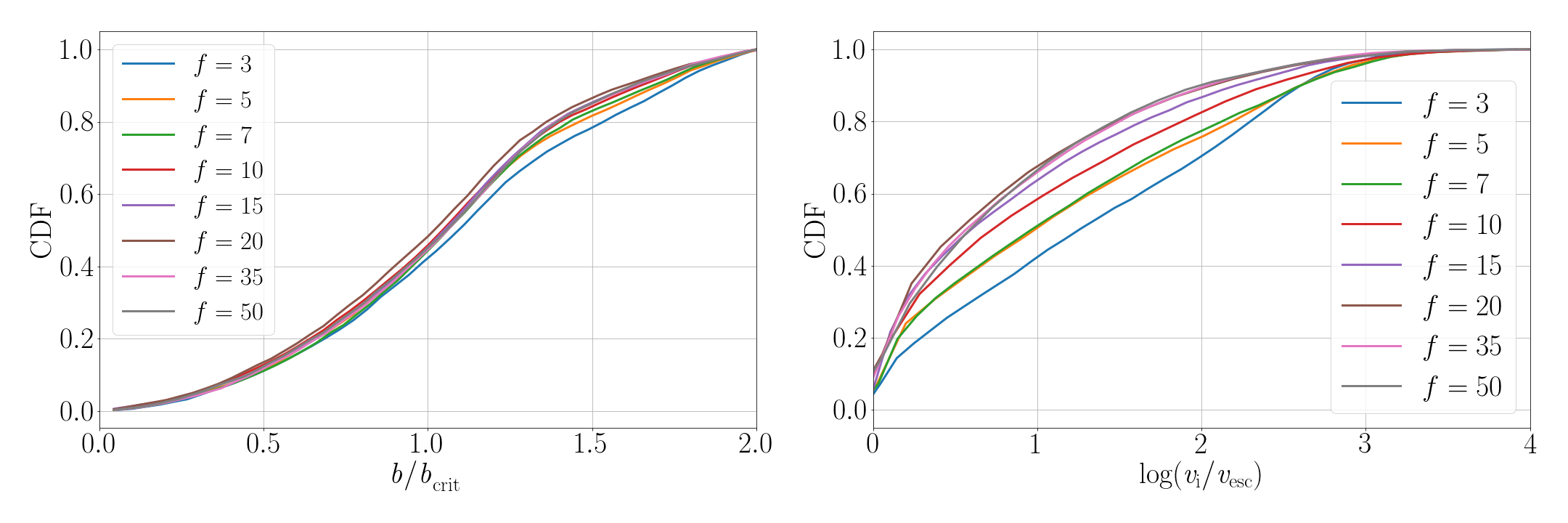}
    \caption{CDF's of $b/b_{\rm crit}$ (left panel) and log$(v_{i}/v_{\rm esc})$ (right panel) of all collisions in systems with different expansion factors.}
    \label{fig:ef_parameter_comparison}
\end{figure*}

Figure \ref{fig:ef_parameter_comparison} shows the CDF for $b/b_{\rm crit}$ and $log(v_{\rm i}/v_{\rm esc})$ for all of the collisions in the runs with a given $f$.  As $f$ increases, impact parameters and impact velocities shift to lower values.  As the radius of the bodies increase, the collision time decreases which results in lower impact velocities.  Collisions with a lower impact velocity are more likely to result in accretion rather than erosion.  This indicates that bodies with larger $f$ are more likely to merge and on a shorter timescale, which rapidly builds more massive planets.  The lower values of $b/b_{\rm crit}$ can be attributed to the fact that the rate of mergers increases with $f$ which spurs runaway growth early on.  As a result, the mass fraction $M_{\rm p}/M_{\rm t}$ decreases quickly with large $f$ and leads to collisions with lower values of $b/b_{\rm crit}$.  We also check the distributions of $b$ as a function of $f$ and find that the distributions of $b$ shifts \textit{slightly} towards lower values with increasing $f$.  If the vertical dispersion given by the mutual inclination ($i_{\rm p}*a_{\rm p}$) isn't much larger than the body sizes with the expansion factor, then the distribution of $b$ will shift towards lower values \citep{Leleu2018}.  Because the change in $b$ with $f$ is so small we conclude that the vertical dispersion is greater than the largest particle radii we considered.

\subsection{Effects on formation timescales}
\begin{figure*}
	\includegraphics[width=2\columnwidth]{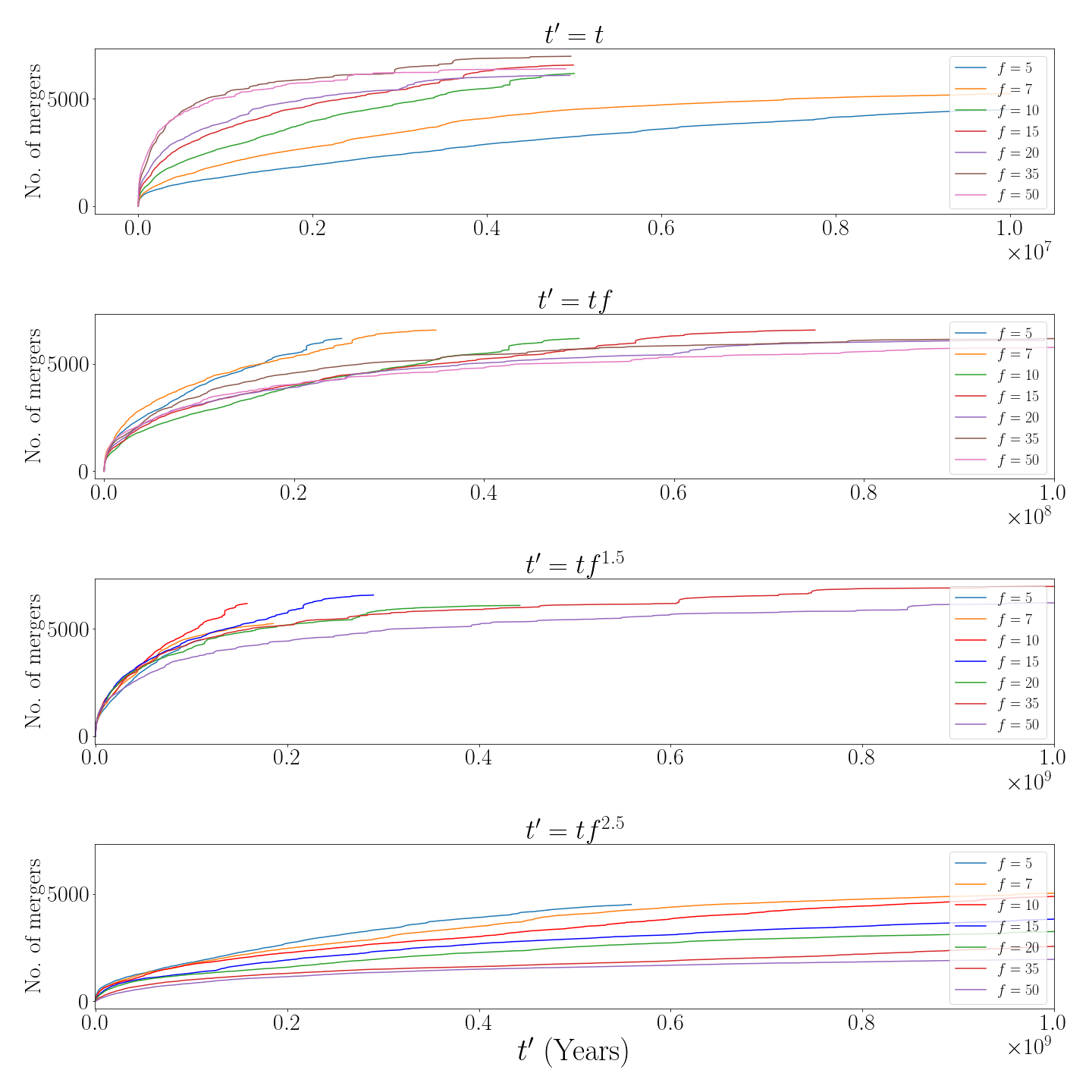}
    \caption{We plot the total number of mergers vs time for each $f$ to understand how the simulation time $t$ corresponds to the effective time of planet formation $t'$ when fragmentation is modeled.  We find that $t'=ft^{1.5}$ is the best fit for approximating how the simulation time scales with the effective time of planet formation for a given $f$.}
    \label{fig:ef_mergers}
\end{figure*}

$f$ and the \textit{effective time}, $t'$, for planet formation are inversely related -- a greater expansion factor will speed up planet formation by increasing the collision rate.  \citet{Childs2021} found that $t' \propto tf^{2.5}$ when only perfect merging is used, but this scaling may not apply to systems that model fragmentation as fragmentation has been shown to significantly alter formation timescales.  Following their methods, we use the total number of mergers as a proxy for how far along planet formation is.  Figure \ref{fig:ef_mergers} shows the number of mergers versus $t'$ for different functions of the simulation time $t$.  We find that $t'=tf^{1.5}$ is the best scaling for approximating the effective time with fragmentation.  $t'$ with fragmentation does not decrease as quickly with increasing $f$ as it does with only perfect merging.  This is expected since the inclusion of disruptive events via fragmentation lengthens planet formation timescales in a non-linear way.

Another property we consider is the relative order in which the terrestrial planets form.  To understand how $f$ affects the order in which the terrestrial planets form we track the time and radii when a body reaches $m=0.2 \, M_{\oplus}$ in one run for each value of $f$.  We find no correlation between $f$ and the order in which the planets form (from the inside-out versus from the outside-in).  The order in which planets form in a run varies, regardless of the value of $f$.  This suggests that the order in which terrestrial planets form is very sensitive to the initial conditions.

\subsection{Correction terms} \label{sec:Correction_term}

To correct for the differences introduced by larger expansion factors, we experiment with different \textit{correction terms} in an effort to replicate the same collision history and final planetary systems as the $f=3$ system.  We experiment with 30 different correction terms in systems with large values of $f$.   We find that each correction term contributes its own effects to the system evolution that complicates the interpretation of the simulation results and do not pursue this approach further.  However, in this section we briefly describe the approaches we explored to correct for the high rate of mergers, low number of fragments, and large semi-major axes that arise in simulations with large values of $f$. 

In order to correct for the faster rate of merging associated with a larger $f$ we experiment with overestimating the impact velocity by various factors of $f$ when resolving the collision.  This was done because when the artificially inflated planets collide, they aren't moving as fast as they would be if they were more dense.  So, we are trying to recreate the collisions as they would occur in real life.  A larger impact velocity $v_{\rm i}$ will produce more erosion events which will inhibit the orbital damping and lower the rate of mergers.  This correction term is used to help offset the fewer number of fragments produced in systems with a larger $f$.

We find that while overestimating $v_{\rm i}$ in our collision routine results in a collision history that is more similar to the $f=3$ case, the systems still retain too much material. This results in planets that are too massive and planets that are at larger semi-major axes.  

We also experiment with expansion factors that scale with the initial semi-major axis of the bodies and with a forced ejection criteria (FEC).  Using $f \propto a$ will reduce the number of mergers as the bodies found at shorter orbital periods will have a smaller $f$, thus maintaining a more constant collision timescale at each orbit. 


The FEC is also a function of semi-major axis which preferentially ejects from the outer regions of the disc if a specific criteria is met during a collision. This is to help correct for planet semi-major axes that increase with $f$.  We find that using $f \propto a$ and the FEC together best reproduces the $f=3$ systems however, the planets preferentially form from the outside-in and still form at larger semi-major axes.

Of the 30 different correction terms we experimented with, we find that each term introduces effects that differentiate the system from the $f=3$ results.  We conclude that studies are better off employing an expansion factor and being aware of the effects from such, as discussed previously, and do not implement any corrections into our fragmentation code to account for the effects of an expansion factor.

\section{Composition tracking}
In addition to our fragmentation code, we developed a post-processing code that tracks how the bulk composition of the bodies change during a collision.  To do so, we make the simplifying assumption that the composition of all the bodies is homogeneous and changes only as a function of mass exchange with bodies of a differing composition. This assumption does not allow for the tracking of core and mantle mass evolution. While we acknowledge that there are many physical processes that occur during planet formation which will alter the composition of the bodies (volatile depletion, fragmentation of differentiated bodies, outgasing, photoevaporation, etc.) and that significant fractions of the volatile budget may lay on the surface of the planet \citep{Burger2020}, our code is a reasonable first approximation for constraining the bulk composition and the maximum relative abundance of a given specie when using realistic starting conditions.  A more nuanced planetary model is beyond our present scope.
\begin{table*}
\centering
\caption{We list the expansion factor $f$ and the average and standard deviations of the WMF of all the planets across all runs for a given $f$.  We also list the average and standard deviations of the planet WMFs in a specified radial range.    These statistics only consider bodies with a mass larger or equal to $0.2 \, M_{\oplus}$ and are not mass averaged.}
\begin{tabular}
{%
  S[table-format=1.0]
  S[table-format=1.2]
  *{2}{ 
    *{2}{S[table-format=1.3(3)]} 
    S[table-format=1.3] 
     S[table-format=1.2]
  }
}
  \toprule
    {$f$} & {WMF} & {$a_{\rm p}/\rm au < 0.8$} & {$0.8 \leq a_{\rm p}/\rm au \leq 1.5$} & {$1.5 < a_{\rm p}/\rm au $}\\
    \cmidrule(lr){1-5}   
    {$3$}&{$0.005 \pm 0.010$}& {$0.001 \pm 0.001$} & {$0.003 \pm 0.003$} & {$0.025 \pm 0.022$} \\
    {$5$}&{$0.006 \pm 0.011$}& {$0.001 \pm 0.001$} & {$0.002 \pm 0.001$} & {$0.015 \pm 0.017$}\\
    {$7$}&{$0.008 \pm 0.013$}& {$0.002 \pm 0.002$} & {$0.003 \pm 0.002$} & {$0.020 \pm 0.018$}\\
    {$10$}&{$0.015 \pm 0.019$}& {$0.004 \pm 0.003$} & {$0.008 \pm 0.004$} & {$0.036 \pm 0.025$}\\
    {$15$}&{$0.009 \pm 0.013$}& {$0.001 \pm 0.002$} & {$0.004 \pm 0.003$} & {$0.020 \pm 0.016$}\\
    {$20$}&{$0.009 \pm 0.015$}& {$0.001 \pm 0.001$} & {$0.002 \pm 0.002$}& {$0.023 \pm 0.017$} \\
    {$35$}&{$0.012 \pm 0.017$}& {$0.001 \pm 0.001$} & {$0.001 \pm 0.002$} & {$0.025 \pm 0.017$}\\
    {$50$}&{$0.013 \pm 0.019$}& {$0.000 \pm 0.000$} & {$0.001 \pm 0.001$} & {$0.026 \pm 0.019$}\\

  \bottomrule
\end{tabular}
\label{tab:ef_WMF}
\end{table*}
\subsection{Composition tracking model}
Our post-processing composition tracking model works in tandem with our fragmentation code.   The fragmentation code produces a collision report to be passed into our \textsc{python} script for composition tracking and follows how mass from various bodies is accreted by the proto-planets.  The collision report along with an input file made by the user are the input needed for our composition code.  The user generated input file lists all of the starting bodies with each body's unique identifier (this is called its hash in \textsc{rebound}), the body mass, and the body's relative abundance for each specie.  The output file returns the last recorded mass and composition for all of the bodies in the system in the same format as the user generated input file.  Details on these formats may be found in the documentation for our code.

To track the composition evolution of the bodies involved in a collision we consider a target with mass $M_{\rm t}$ and initial composition 
\begin{equation}
    \mathbf{X} = \sum_{i=1}^{n} x_{i}=1,
\end{equation}
and a projectile with mass $M_{\rm p}$ and initial composition 
\begin{equation}
    \mathbf{Y} = \sum_{i=1}^{n} y_{i}=1,
\end{equation}
for $n$ number of species being tracked, and $x_i$, $y_i$ being the relative abundance of the $i^{th}$ specie for the target and projectile respectively. 

If the collision results in an elastic bounce with no mass exchange, then the composition of each body remains the same.  If the collision results in a merger or partial accretion of the projectile, the composition of the target becomes
\begin{equation}
    \mathbf{X'}= \frac{1}{M_{\rm t} + M_{\rm p}'} \left [ M_{\rm t} \sum_{i=1}^{n} x_{i} + M_{\rm p}'/M_{\rm t} \sum_{i=1}^{n} y_{i} \right ].
\end{equation}
where $M_{\rm p}'$ is the mass of the projectile that is accreted by the target.  If any fragments are produced from the projectile, they are assigned the composition of the projectile.

If the target becomes eroded, then the composition of the target remains the same and it is assigned the mass of the largest remnant, $M_{\rm lr}$. The composition of the new fragment(s) becomes
\begin{equation}
    \mathbf{Z'}=\frac{1}{M_{\rm tot}-M_{\rm lr}} \left [   (M_{\rm t} -M_{\rm lr}) \sum_{i=1}^{n} x_{i} + M_{\rm p}/M_{\rm t} \sum_{i=1}^{n} y_{i}   \right ].
\end{equation}
If multiple fragments are produced in a collision, they all receive the same composition.

\begin{table*}
\centering
\caption{Collision model used, simulation time and the average values and standard deviations for the final terrestrial planet multiplicity, planet mass ($M_{\rm p}$), semi-major axis ($a_{\rm p}$), eccentricity ($e$), and inclination ($i$).  We also list the WMF for all the planets in the WMF column, and the WMF of planets found in a specified radial range.  These statistics only consider bodies with a mass larger or equal to $0.2 \, M_{\oplus}$ and are not mass averaged.} 
\resizebox{\linewidth}{!}{\begin{tabular}{cccccccccc}
  \toprule
    {Model} &{Multiplicity} & {$M_{\rm p}/M_\oplus$} & {$a_{\rm p}/ \rm au$} & {$e_{\rm p}$} & {$i^{\circ}_{\rm p}$} & {WMF}& {$a_{\rm p}/\rm au < 0.8$} & {$0.8 \leq a_{\rm p}/\rm au \leq 1.5$} & {$1.5 < a_{\rm p}/\rm au $}\\
    \cmidrule(lr){1-10}   
    {Fragmentation} & {$3.4 \pm 0.8$} & {$0.98 \pm 0.70$} & {$1.24 \pm 0.67$} & {$0.08 \pm 0.05$} & {$2.88 \pm 2.46$} & {$0.015 \pm 0.019$}& {$0.004 \pm 0.003$}& {$0.008 \pm 0.004$}& {$0.036 \pm 0.025$}\\
    
    {Perfect Merging}& {$2.9 \pm 1.04$} & {$1.21 \pm 0.81$} & {$1.34 \pm 0.68$} & {$0.12 \pm 0.10$} & {$3.60 \pm 3.01$}& {$0.009 \pm 0.013$}& {$0.001 \pm 0.001$}& {$0.005 \pm 0.005$}& {$0.019 \pm 0.016$}\\

  \bottomrule
\end{tabular}}
\label{tab:frag_merge_systems}
\end{table*}

\section{Volatile delivery}
Using our fragmentation and composition tracking code we study how expansion factors and fragmentation affect the delivery of water to the inner terrestrial region.  Using the same disc from Section \ref{sec:n_body_setup} we apply an initial water content distribution to each body as is used in \citet{Raymond2006}.
The initial water content is designed to reproduce the water content of chondritic classes of meteorites.  It is a discrete partition of water abundance as follows: inside $2 \, \rm au$ bodies are initially dry, bodies between $2 \, \rm and \, 2.5 \, \rm au$ contain $0.1\%$ water by mass, and outside of $2.5 \, \rm au$ bodies have an initial water content of $5\%$ by mass.  The top panel of Figure \ref{fig:WMF} shows the initial water mass fraction (WMF) for the disc.  After assigning each body an initial water content, we use our composition tracking code to follow the evolution of the WMF for the planets in the systems.

\subsection{Expansion factor effects on water delivery}
Table \ref{tab:ef_WMF} lists the average and standard deviation of the planet WMFs that formed across all ten runs for a given $f$. We evaluate all the systems at the times listed in Table \ref{tab:ef_collision_history}.  Following \citet{Raymond2006} we define the habitable zone around a Sun-like star as $0.8-1.5 \, \rm au$ and now define a planet as a body having a mass $> 0.2 \, M_{\oplus}$.  We also list the average planet WMF interior to the habitable zone, inside the habitable zone, and exterior to the habitable zone.

In general, the average WMF of the planets increases with $f$.  The only exception to this trend is $f=10$, which is an outlier.  For all $f$, planet WMF increases with $a_{\rm p}$ and the water gradient steepens as $f$ increases.  Larger expansion factors impede radial mixing as particles with larger radii are more likely to result in a collision than a close encounter which leads to bodies on less eccentric orbits. The collisions are also detected earlier on, before the bodies have time to move away from their original location.  The WMF for Earth is 0.001.  As expected, the $f=3$ system produces planets with WMFs most similar to Earth's.


\subsection{Fragmentation effects on water delivery}
\begin{figure*}
	\includegraphics[width=1.5\columnwidth, height=.5\textheight]{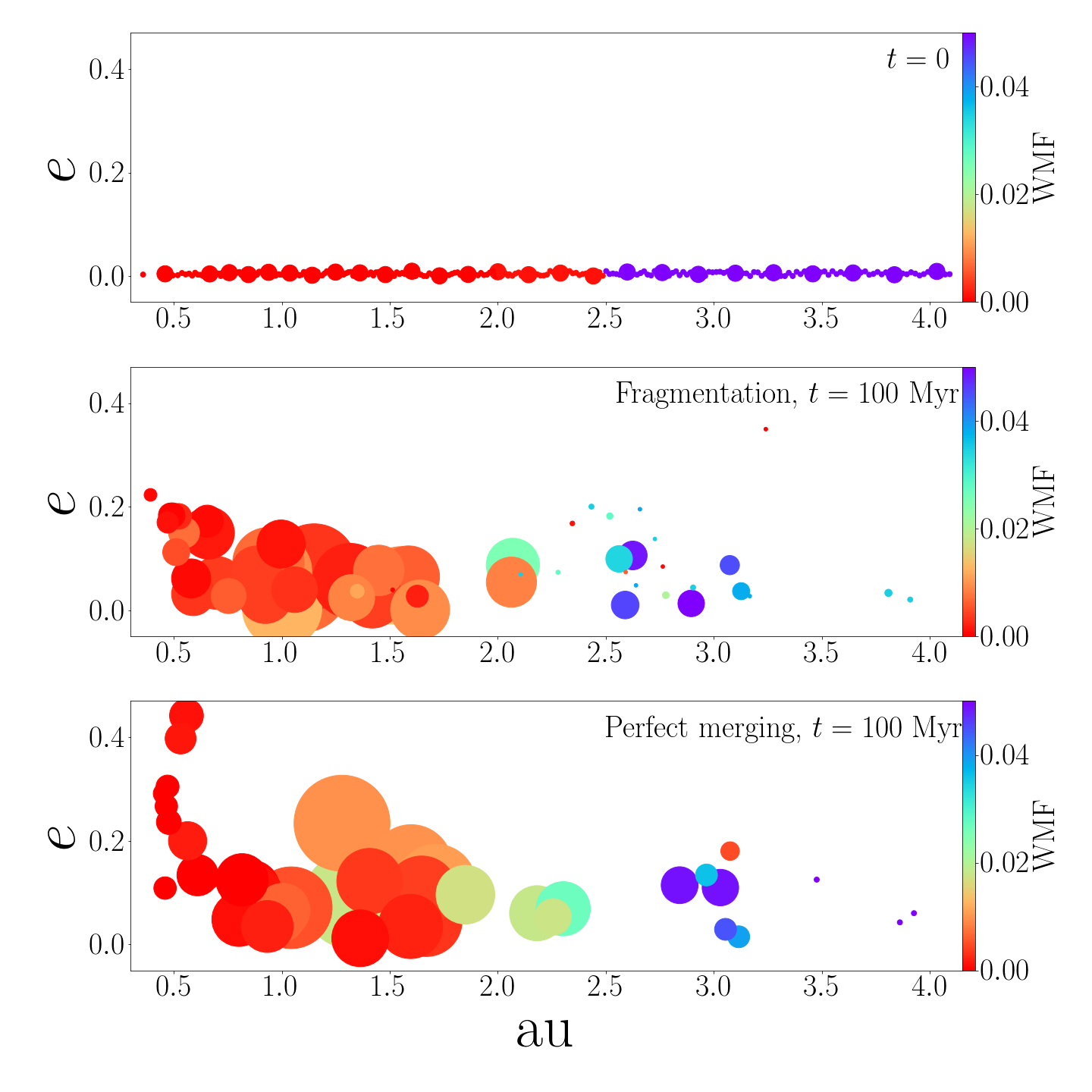}
    \caption{Eccentricity vs. semi-major axis for all the bodies in our simulation at $t=0$ and for the remaining bodies in all simulations after $100 \, \rm Myr$.  The size of the marker is proportional to the mass of the body and the colour of the marker represents the water mass fraction (WMF) as listed on the colour bars.  The top panel shows the initial distribution of the bodies and they're starting water content.  The middle panel shows the final planetary systems from all ten runs after integrating for $100 \, \rm Myr$ with fragmentation, and the final planetary systems from all ten runs with only perfect merging in the bottom panel.}
    \label{fig:WMF}
\end{figure*}

We first consider how fragmentation affects the final planetary system.  We perform ten runs with fragmentation and ten runs with only perfect merging using $f=10$ to reduce the computation time.  We integrate the systems for $100 \, \rm Myr$.  The first two columns in Table \ref{tab:frag_merge_systems} list the average planet multiplicity and average planet mass, semi-major axis, eccentricity and inclination per run for systems with fragmentation and systems with only perfect merging.  We find that fragmentation produces systems with similar planet multiplicities but planets with lower mass than the systems with only perfect merging.  The planet orbits have slightly lower semi-major axis, eccentricity, and inclination with fragmentation, which is likely the result of dynamical friction from the fragments in the system.

Figure \ref{fig:WMF} shows the results from simulations with fragmentation (middle panel) and with only perfect merging (bottom panel) after $10 \, \rm Myr$ of integration time.  We plot eccentricity vs. semi-major axis for the remaining bodies in all the simulations.  The area of the marker is proportional to the mass of the body and the colour of the marker represents the water mass fraction (WMF) as listed on the colour bars.  Table \ref{tab:frag_merge_systems} lists the average WMF of a planet for systems with each collision model and the WMF of the planets found interior to the habitable zone, inside the habitable zone, and exterior to the habitable zone.  We find that on average, fragmentation produces planets with a higher WMF than the planets formed via perfect-merging.  We attribute this to a more enhanced compositional mixing from the creation and accretion of fragments.  Fragments produced by bodies further out and with a larger WMF are able to be accreted by closer-in planets, increasing the average planet WMF. This is also reflected in the WMF of the radial bins we consider.  In all three radial bins, fragmentation produces planets with larger WMFs than perfect merging.  The differences in WMFs between fragmentation and merging are most pronounced in the inner regions where fragmentation enables water delivery to the inner regions and perfect merging does not.  However, perfect merging produces planets with a high WMF between $~1.3-2.3 \, \rm au$.  We find four planets with a WMF of $2\%$ in this range.  The lack of radial mixing of material that arises from assuming perfect merging leads to an overabundance of dry planets but can also produce very wet planets.  If perfect merging does add water to a body, it is more likely to do so in large increments whereas fragmentation allows for water delivery to happen in smaller increments.

During Earth's formation, a significant amount of water likely evaporated off during the giant impact phase \citep{Lock2017}.  With respect to Earth, perfect merging and fragmentation overestimate the WMF by factors of 9 and 15, respectively. However, our simulations consider planets formed out to $4 \, \rm au$, volatile loss is not accounted for, and we use thee expansion factor $f=10$.  

In addition to forming more water rich planets, fragmentation allows for mass exchange and this mixing of bodies leads to systems with lower gradients of water, and water is found over a larger range of radii than systems that assume only perfect merging.  \citet{Emsenhuber2020} also developed their own fragmentation model using machine learning methods and a composition tracking code.  Although the specifics of their models differ, they also found that fragmentation leads to more compositional mixing throughout planet formation.  In general, as is expected with in-situ formation, we observe that the final location of a planet is correlated with its final WMF as planets found at smaller radii will have lower WMF than planets formed at larger radii.

\section{Conclusions}

We presented a fragmentation module and a composition tracking post-processing code to be used with the $n$-body code \textsc{rebound}.  Our fragmentation code allows for improved studies of planet formation by modeling more realistic collisions during the late stage of planet formation.  We compared our fragmentation results to the fragmentation results of the $n$-body code \textsc{mercury} by \citet{Chambers2013} and found good agreement in collision outcomes for the same impact velocity-impact parameter phase space, and the overall evolution of the system.

In our $n$-body simulations with fragmentation, we inflated the particle radii by an expansion factor $f$ and experimented with various values of $f$ to understand how they affect the collision history and final planetary system.  We found that as the expansion factor increases, so do the rate of mergers, which leads to planetary systems with more planets and planets found at larger radii.  We experimented with various correction terms in an effort to offset the errors associated with the expansion factors. No correction terms were found to compensate the effects of expansion factors without introducing their own error and we conclude that, until a reliable correction can be identified, when users apply an expansion factor, they recognize the systematic errors associated with their use.

Our composition tracking code is a post-processing \textsc{Python} code which uses the output from our fragmentation module to track how the composition of bodies changes as a function of mass exchange.  Our code assumes that all the bodies are non-differentiated with a homogeneous composition.  We studied how expansion factors and fragmentation affect volatile delivery into the inner disc regions using our composition tracking code.  Using an initial water distribution adopted from \citet{Raymond2006}, we track how the WMF of the planets evolves in systems with various values of $f$. We find that on average, as $f$ increases so does the average WMF of the planets.  Radial mixing decreases with increasing $f$ as collisions happen early on, before the bodies have time to grow to excited orbits and move away from their original location.  We also compared the resulting WMF in systems that use the same initial conditions but model either fragmentation or only perfect merging.  We found that on average, accounting for fragmentation produces planets with a higher WMF and that fragmentation allows for more radial mixing.  The exchange of material between bodies of differing WMF produces systems with a shallower gradient of WMF over the terrestrial radial range of the disc.


We make both our fragmentation module and composition tracking code open-source to be used with \textsc{rebound} for future studies.


\section*{Acknowledgements}

We thank an anonymous referee for useful comments that improved the manuscript.  We thank Hanno Rein for helpful discussions. We thank John Chambers for making his code available for comparison.  Computer support was provided by UNLV’s National Supercomputing Center. AC acknowledges support from a UNLV graduate assistantship.  AC and JS acknowledges support from the NSF through grant AST-1910955.  Simulations in this paper made use of the rebound code which can be downloaded freely at \url{http://github.com/hannorein/rebound}.

\section*{Data Availability}\label{sec:Data}

The fragmentation and composition tracking codes for \textsc{rebound} with documentation, can be found at \url{http://github.com/ANNACRNN/REBOUND_fragmentation}. The $n$-body simulation results can be reproduced with the {\sc rebound} code (Astrophysics Source Code Library identifier {\tt ascl.net/1110.016}).  The data underlying this article will be shared on reasonable request to the corresponding author.



\bibliographystyle{mnras}
\bibliography{main} 

\begin{thebibliography}{}
\makeatletter
\relax
\def\mn@urlcharsother{\let\do\@makeother \do\$\do\&\do\#\do\^\do\_\do\%\do\~}
\def\mn@doi{\begingroup\mn@urlcharsother \@ifnextchar [ {\mn@doi@}
  {\mn@doi@[]}}
\def\mn@doi@[#1]#2{\def\@tempa{#1}\ifx\@tempa\@empty \href
  {http://dx.doi.org/#2} {doi:#2}\else \href {http://dx.doi.org/#2} {#1}\fi
  \endgroup}
\def\mn@eprint#1#2{\mn@eprint@#1:#2::\@nil}
\def\mn@eprint@arXiv#1{\href {http://arxiv.org/abs/#1} {{\tt arXiv:#1}}}
\def\mn@eprint@dblp#1{\href {http://dblp.uni-trier.de/rec/bibtex/#1.xml}
  {dblp:#1}}
\def\mn@eprint@#1:#2:#3:#4\@nil{\def\@tempa {#1}\def\@tempb {#2}\def\@tempc
  {#3}\ifx \@tempc \@empty \let \@tempc \@tempb \let \@tempb \@tempa \fi \ifx
  \@tempb \@empty \def\@tempb {arXiv}\fi \@ifundefined
  {mn@eprint@\@tempb}{\@tempb:\@tempc}{\expandafter \expandafter \csname
  mn@eprint@\@tempb\endcsname \expandafter{\@tempc}}}

\bibitem[\protect\citeauthoryear{Agnor, Canup  \& Levison}{Agnor
  et~al.}{1999}]{AGNOR1999}
Agnor C.~B.,  Canup R.~M.,   Levison H.~F.,  1999, Icarus, 142, 219

\bibitem[\protect\citeauthoryear{Asphaug}{Asphaug}{2010}]{Asphaug2010}
Asphaug E.,  2010, Geochemistry, 70, 199

\bibitem[\protect\citeauthoryear{{Bonsor}, {Leinhardt}, {Carter}, {Elliott},
  {Walter}  \& {Stewart}}{{Bonsor} et~al.}{2015}]{Bonsor2015}
{Bonsor} A.,  {Leinhardt} Z.~M.,  {Carter} P.~J.,  {Elliott} T.,  {Walter}
  M.~J.,   {Stewart} S.~T.,  2015, \mn@doi [\icarus]
  {10.1016/j.icarus.2014.10.019}, \href
  {https://ui.adsabs.harvard.edu/abs/2015Icar..247..291B} {247, 291}

\bibitem[\protect\citeauthoryear{{Burger}, {Bazs{\'o}}  \&
  {Sch{\"a}fer}}{{Burger} et~al.}{2020}]{Burger2020}
{Burger} C.,  {Bazs{\'o}} {\'A}.,   {Sch{\"a}fer} C.~M.,  2020, \mn@doi [\aap]
  {10.1051/0004-6361/201936366}, \href
  {https://ui.adsabs.harvard.edu/abs/2020A&A...634A..76B} {634, A76}

\bibitem[\protect\citeauthoryear{Carter, Leinhardt, Elliott, Walter  \&
  Stewart}{Carter et~al.}{2015}]{Carter2015}
Carter P.~J.,  Leinhardt Z.~M.,  Elliott T.,  Walter M.~J.,   Stewart S.~T.,
  2015, The Astrophysical Journal, 813, 72

\bibitem[\protect\citeauthoryear{{Chambers}}{{Chambers}}{1999}]{Chambers1999}
{Chambers} J.~E.,  1999, \mn@doi [\mnras] {10.1046/j.1365-8711.1999.02379.x},
  \href {https://ui.adsabs.harvard.edu/abs/1999MNRAS.304..793C} {304, 793}

\bibitem[\protect\citeauthoryear{Chambers}{Chambers}{2001a}]{CHAMBERS2001}
Chambers J.,  2001a, Icarus, 152, 205

\bibitem[\protect\citeauthoryear{Chambers}{Chambers}{2001b}]{CHAMBERS2010}
Chambers J.,  2001b, Icarus, 152, 205

\bibitem[\protect\citeauthoryear{{Chambers}}{{Chambers}}{2013}]{Chambers2013}
{Chambers} J.~E.,  2013, \mn@doi [\icarus] {10.1016/j.icarus.2013.02.015},
  \href {https://ui.adsabs.harvard.edu/abs/2013Icar..224...43C} {224, 43}

\bibitem[\protect\citeauthoryear{Chambers \& Wetherill}{Chambers \&
  Wetherill}{1998}]{CHAMBERS1998}
Chambers J.,  Wetherill G.,  1998, Icarus, 136, 304

\bibitem[\protect\citeauthoryear{{Childs} \& {Martin}}{{Childs} \&
  {Martin}}{2021}]{Childs2021}
{Childs} A.~C.,  {Martin} R.~G.,  2021, \mn@doi [\mnras]
  {10.1093/mnras/stab2419}, \href
  {https://ui.adsabs.harvard.edu/abs/2021MNRAS.507.3461C} {507, 3461}

\bibitem[\protect\citeauthoryear{Childs, Quintana, Barclay  \& Steffen}{Childs
  et~al.}{2019}]{Childs_2019}
Childs A.~C.,  Quintana E.,  Barclay T.,   Steffen J.~H.,  2019, \mn@doi
  [Monthly Notices of the Royal Astronomical Society] {10.1093/mnras/stz385},
  485, 541–549

\bibitem[\protect\citeauthoryear{{Emsenhuber}, {Cambioni}, {Asphaug},
  {Gabriel}, {Schwartz}  \& {Furfaro}}{{Emsenhuber}
  et~al.}{2020}]{Emsenhuber2020}
{Emsenhuber} A.,  {Cambioni} S.,  {Asphaug} E.,  {Gabriel} T. S.~J.,
  {Schwartz} S.~R.,   {Furfaro} R.,  2020, \mn@doi [\apj]
  {10.3847/1538-4357/ab6de5}, \href
  {https://ui.adsabs.harvard.edu/abs/2020ApJ...891....6E} {891, 6}

\bibitem[\protect\citeauthoryear{Gabriel, Jackson, Asphaug, Reufer, Jutzi  \&
  Benz}{Gabriel et~al.}{2020}]{Gabriel_2020}
Gabriel T. S.~J.,  Jackson A.~P.,  Asphaug E.,  Reufer A.,  Jutzi M.,   Benz
  W.,  2020, The Astrophysical Journal, 892, 40

\bibitem[\protect\citeauthoryear{{Genda}, {Kokubo}  \& {Ida}}{{Genda}
  et~al.}{2012}]{Genda2012}
{Genda} H.,  {Kokubo} E.,   {Ida} S.,  2012, \mn@doi [\apj]
  {10.1088/0004-637X/744/2/137}, \href
  {https://ui.adsabs.harvard.edu/abs/2012ApJ...744..137G} {744, 137}

\bibitem[\protect\citeauthoryear{Kokubo \& Genda}{Kokubo \&
  Genda}{2010}]{Kokubo_2010}
Kokubo E.,  Genda H.,  2010, The Astrophysical Journal, 714, L21

\bibitem[\protect\citeauthoryear{Kokubo \& Ida}{Kokubo \&
  Ida}{1996}]{Kokubo_1996}
Kokubo E.,  Ida S.,  1996, Icarus, 123, 180

\bibitem[\protect\citeauthoryear{Kokubo \& Ida}{Kokubo \&
  Ida}{2000}]{KOKUBO2000}
Kokubo E.,  Ida S.,  2000, Icarus, 143, 15

\bibitem[\protect\citeauthoryear{Kokubo \& Ida}{Kokubo \&
  Ida}{2002}]{Kokubo_2002}
Kokubo E.,  Ida S.,  2002, The Astrophysical Journal, 581, 666

\bibitem[\protect\citeauthoryear{L{\'e}cuyer, Gillet  \& Robert}{L{\'e}cuyer
  et~al.}{1998}]{LECUYER1998}
L{\'e}cuyer C.,  Gillet P.,   Robert F.,  1998, Chemical Geology, 145, 249

\bibitem[\protect\citeauthoryear{{Leinhardt} \& {Richardson}}{{Leinhardt} \&
  {Richardson}}{2005}]{Leinhardt2005}
{Leinhardt} Z.~M.,  {Richardson} D.~C.,  2005, \mn@doi [\apj] {10.1086/429402},
  \href {https://ui.adsabs.harvard.edu/abs/2005ApJ...625..427L} {625, 427}

\bibitem[\protect\citeauthoryear{Leinhardt \& Stewart}{Leinhardt \&
  Stewart}{2011a}]{Leinhardt_2011}
Leinhardt Z.~M.,  Stewart S.~T.,  2011a, The Astrophysical Journal, 745, 79

\bibitem[\protect\citeauthoryear{Leinhardt \& Stewart}{Leinhardt \&
  Stewart}{2011b}]{Leinhardt_2011b}
Leinhardt Z.~M.,  Stewart S.~T.,  2011b, \mn@doi [The Astrophysical Journal]
  {10.1088/0004-637x/745/1/79}, 745, 79

\bibitem[\protect\citeauthoryear{{Leleu}, {Jutzi}  \& {Rubin}}{{Leleu}
  et~al.}{2018}]{Leleu2018}
{Leleu} A.,  {Jutzi} M.,   {Rubin} M.,  2018, \mn@doi [Nature Astronomy]
  {10.1038/s41550-018-0471-7}, \href
  {https://ui.adsabs.harvard.edu/abs/2018NatAs...2..555L} {2, 555}

\bibitem[\protect\citeauthoryear{Lock \& Stewart}{Lock \&
  Stewart}{2017}]{Lock2017}
Lock S.~J.,  Stewart S.~T.,  2017, Journal of Geophysical Research: Planets,
  122, 950

\bibitem[\protect\citeauthoryear{Marty}{Marty}{2012}]{Marty2012}
Marty B.,  2012, \mn@doi [Earth and Planetary Science Letters]
  {10.1016/j.epsl.2011.10.040}, 313-314, 56

\bibitem[\protect\citeauthoryear{Morbidelli, Chambers, Lunine, Petit, Robert,
  Valsecchi  \& Cyr}{Morbidelli et~al.}{2000}]{Morbidelli2000}
Morbidelli A.,  Chambers J.,  Lunine J.~I.,  Petit J.~M.,  Robert F.,
  Valsecchi G.~B.,   Cyr K.~E.,  2000, Meteoritics \& Planetary Science, 35,
  1309

\bibitem[\protect\citeauthoryear{Moriarty, Madhusudhan  \& Fischer}{Moriarty
  et~al.}{2014}]{Moriarty2014}
Moriarty J.,  Madhusudhan N.,   Fischer D.,  2014, 787, 81

\bibitem[\protect\citeauthoryear{{Morishima}, {Stadel}  \& {Moore}}{{Morishima}
  et~al.}{2010}]{Morishima2010}
{Morishima} R.,  {Stadel} J.,   {Moore} B.,  2010, \mn@doi [\icarus]
  {10.1016/j.icarus.2009.11.038}, \href
  {https://ui.adsabs.harvard.edu/abs/2010Icar..207..517M} {207, 517}

\bibitem[\protect\citeauthoryear{{O'Brien}, {Morbidelli}  \&
  {Levison}}{{O'Brien} et~al.}{2006}]{Obrien2006}
{O'Brien} D.~P.,  {Morbidelli} A.,   {Levison} H.~F.,  2006, \mn@doi [\icarus]
  {10.1016/j.icarus.2006.04.005}, \href
  {https://ui.adsabs.harvard.edu/abs/2006Icar..184...39O} {184, 39}

\bibitem[\protect\citeauthoryear{Quintana, Barclay, Borucki, Rowe  \&
  Chambers}{Quintana et~al.}{2016}]{Quintana_2016}
Quintana E.~V.,  Barclay T.,  Borucki W.~J.,  Rowe J.~F.,   Chambers J.~E.,
  2016, \mn@doi [The Astrophysical Journal] {10.3847/0004-637x/821/2/126}, 821,
  126

\bibitem[\protect\citeauthoryear{Rafikov}{Rafikov}{2003}]{Rafikov_2003}
Rafikov R.~R.,  2003, The Astronomical Journal, 125, 942

\bibitem[\protect\citeauthoryear{{Raymond}, {Quinn}  \& {Lunine}}{{Raymond}
  et~al.}{2004}]{Raymond2004}
{Raymond} S.~N.,  {Quinn} T.,   {Lunine} J.~I.,  2004, \mn@doi [\icarus]
  {10.1016/j.icarus.2003.11.019}, \href
  {https://ui.adsabs.harvard.edu/abs/2004Icar..168....1R} {168, 1}

\bibitem[\protect\citeauthoryear{{Raymond}, {Quinn}  \& {Lunine}}{{Raymond}
  et~al.}{2006}]{Raymond2006}
{Raymond} S.~N.,  {Quinn} T.,   {Lunine} J.~I.,  2006, \mn@doi [\icarus]
  {10.1016/j.icarus.2006.03.011}, \href
  {https://ui.adsabs.harvard.edu/abs/2006Icar..183..265R} {183, 265}

\bibitem[\protect\citeauthoryear{{Raymond}, {O'Brien}, {Morbidelli}  \&
  {Kaib}}{{Raymond} et~al.}{2009}]{Raymond2009}
{Raymond} S.~N.,  {O'Brien} D.~P.,  {Morbidelli} A.,   {Kaib} N.~A.,  2009,
  \mn@doi [\icarus] {10.1016/j.icarus.2009.05.016}, \href
  {https://ui.adsabs.harvard.edu/abs/2009Icar..203..644R} {203, 644}

\bibitem[\protect\citeauthoryear{{Rein} \& {Liu}}{{Rein} \&
  {Liu}}{2012}]{Rein2012}
{Rein} H.,  {Liu} S.~F.,  2012, \mn@doi [\aap] {10.1051/0004-6361/201118085},
  \href {https://ui.adsabs.harvard.edu/abs/2012A&A...537A.128R} {537, A128}

\bibitem[\protect\citeauthoryear{{Rein} et~al.,}{{Rein}
  et~al.}{2019}]{Rein2019}
{Rein} H.,  et~al., 2019, \mn@doi [\mnras] {10.1093/mnras/stz769}, \href
  {https://ui.adsabs.harvard.edu/abs/2019MNRAS.485.5490R} {485, 5490}

\bibitem[\protect\citeauthoryear{Righter \& O{\textquoteright}Brien}{Righter \&
  O{\textquoteright}Brien}{2011}]{Righter11}
Righter K.,  O{\textquoteright}Brien D.~P.,  2011, Proceedings of the National
  Academy of Sciences, 108, 19165

\bibitem[\protect\citeauthoryear{Shoemaker}{Shoemaker}{1961}]{Shoemaker1962}
Shoemaker E.~M.,  1961, in KOPAL Z.,  ed., , Physics and Astronomy of the Moon.
Academic Press, pp 283--359

\bibitem[\protect\citeauthoryear{Stewart \& Leinhardt}{Stewart \&
  Leinhardt}{2009}]{Stewart_2009}
Stewart S.~T.,  Leinhardt Z.~M.,  2009, The Astrophysical Journal, 691, L133

\bibitem[\protect\citeauthoryear{Weidenschilling}{Weidenschilling}{1977}]{Weidenschilling1977}
Weidenschilling S.~J.,  1977, Astrophysics and Space Science, 51, 153

\makeatother
\end{thebibliography}

%

\bsp	
\label{lastpage}
\end{document}